\newcommand{\be}{\begin{eqnarray}}
\newcommand{\ee}{\end{eqnarray}}

\documentclass[12pt]{article}

\usepackage{amsmath,amssymb}
\usepackage{graphicx}
\usepackage{cite}
\usepackage{bm}
\begin{document}

\vskip 2cm

\begin{center}

{\Large Spontaneous generation of the Newton constant in the renormalizable gravity theory} 
\footnote{Preprint ITEP-63 (1982) published in the Proceedings 
 of the conference on Group Theoretic Methods in Physics 
(Zvenigorod, 1982).}

\vskip 4cm

A.V. \textsc{Smilga} \\
\texttt{smilga@subatech.in2p3.fr}

\end{center}

\vskip 5cm

\begin{abstract}

The conformal supergravity is suggested as a realistic 
theory for gravity interactions. It displays the spontaneous breaking 
of the conformal symmetry which results in appearance of the term
 proportional to the scalar curvature $R$ in the effective potential
 with respect to small metric fluctuations.

\end{abstract}

\newpage

The existing gravity theory, Einstein's general relativity,
 is known to include the dimensional coupling constant $k = \frac {8\pi G}{c^2}$; 
$k \sim m_P^{-2}$ in the $\hbar = c = 1$ unit system.
 This theory is beautiful
and consistent at the classical level, but the dimensional
 constant leads to serious troubles in attempts of quantizing, 
the corresponding quantum theory appearing not to be renormalizable.
 A hope exists that a supersymmetric generalization of general 
relativity is renormalizable. But the simplest variant of such a theory 
including one additional Rarita-Schwinger field, seemingly, does not solve the problem
\cite{Peter}. The renormalizability of more complicated theories is
 less investigated. However, even $N=8$ supergravity 
is known to possess nontrivial counterterms arising at the 8-loop level
\cite{Kallosh}, and the renormalizability of such a theory is highly questionable.

There is another known theory with a dimensional constant --- the four-fermion weak interactions
theory due to Fermi which is nonrenormalizable too. The problem has been presently 
solved in this case ---  the Fermi Lagrangian turned out not to be a fundamental one
 but in essence a low energy limit of the renormalizable Weinberg--Salam theory. 
The dimensional constant $
G_F$ arises as a result of the spontaneous breaking of gauge symmetry due to Higgs 
condensate generation. The wish is natural to build the theory where the gravity constant
 $k$ does not appear in the fundamental Lagrangian and  Einstein's action arises as the
 low energy limit of a renormalizable theory.

This idea pronounced first by Sakharov \cite{Sakharov} attracted recently a considerable
attention \cite{Adler,Hasslacher,Zee}. In the simplest model, the additional scalar field $\phi$
is introduced and the initial Lagrangian is chosen in the form 
  \be
{\cal L} \ =\ gR \phi^2 - V(\phi) + \frac 12 (\partial_\mu \phi)^2 \, .
 \ee
If the potential $V(\phi)$ has the minimum at the nonzero $\phi$ values 
(analogous to the Weinberg--Salam model), Einstein's action with a
 dimensional constant is generated in the low energy
limit. However, this model seems not to be very attractive as the additional scalar field
with a condensate of the Plank mass order is required.

The important observation \cite{Adler} is that such a scalar field is not necessary 
at all, and the gravity constant arises spontaneously in almost any theory. 
To see this, write down the action for matter fields in the curved space,
  \be
S \ =\ \int d^4x \, \sqrt{-g} \, {\cal L}^{\rm matt.} 
  \ee
 and find out the effective potential with respect to small fluctuations of the metric.

The theories of Yang--Mills type displaying the spontaneous breaking 
of the scale symmetry are especially attractive. 

In Ref.\cite{Hasslacher}, the effective potential $V(g)$ for the Yang--Mills 
theory was calculated in the dilute instanton gas model. The finite gravity constant 
arises here, indeed. By the order of magnitude, $k \sim \Lambda^{-2}$ where $\Lambda$
 is the fundamental scale parameter of the theory. Thus, to generate the observed Newton
 constant, the Yang-Mills field with the scale parameter of the Plank mass order is
 required (for QCD, $\Lambda
\sim 100 \, {\rm MeV} \ll m_P$).

The main idea of this article is a remark that the introduction of additional matter fields
is unnecessary and the mechanism for generation of the Newton constant in the {\it 
pure gravity theory} can be suggested. Consider the theory with the Lagrangian presenting 
a square of the curvature tensor,

\be
\label{Riemann2}
S \ =\ \frac 1\lambda \int d^4x \, \sqrt{-g} \, R_{\mu\nu\rho\sigma}  R^{\mu\nu\rho\sigma} \, . 
  \ee  
The coupling constant $\lambda$ is dimensionless in this theory so 
that the theory is renormalizable. Moreover, the investigation due 
to Fradkin and Tseytlin \cite{Fradkin} showed that this theory 
(as well as the theories with Lagrangians ${\cal L} \sim 
R_{\mu\nu}^2$ or ${\cal L} \sim R^2$) enjoys the asymptotic freedom.

The choice of action in the form (\ref{Riemann2}) will be discussed a bit later. 
We note in advance that it is not our final choice.

The asymptotic freedom implies that, at very high energies, the constant $\lambda$ 
turns to zero, but there exists a scale (it is natural to assume that it is the Plank scale)
where $\lambda$ is no longer small. In the full analogy with the Yang--Mills theory, 
the scale symmetry breaks spontaneously at Plank distances, the condensates of the kind 
$\langle R_{\mu\nu\rho\sigma}  R^{\mu\nu\rho\sigma} \rangle_0$ and also those including 
 higher powers of the curvature tensor do appear. This results, in particular, 
in the spontaneous
generation of the gravity constant, $k \sim 1/m_P^2$ following the mechanism
investigated in Refs. \cite{Adler,Hasslacher,Zee}.

The basic argument in favor of this phenomenon is the existence of instanton solutions in 
such a theory whose physical meaning is very close to the instanton solutions
in the Yang--Mills theory. The solutions of this kind were found in Ref. \cite{Eguchi}.
The gravitational instanton presents  bent 4-dimensional Euclidean space asymptotically 
flat at infinity at all directions with a self--dual Riemann tensor,
 \be
R_{\mu\nu\rho\sigma}  \ =\ \pm \frac 12 \epsilon_{\mu\nu\kappa\lambda} \, 
R^{\kappa\lambda}_{\ \ \rho\sigma} \ =\ \frac 14  \epsilon_{\mu\nu\kappa\lambda}
  \epsilon_{\rho\sigma\epsilon\delta} R^{\kappa\lambda\epsilon\delta} \, .
 \ee
It can be easily seen that this space satisfied the Einstein's equations $R_{\mu\nu} = 0$
(this follows immediately from the cyclic condition for the Riemann tensor) and also the 
equation of motion corresponding to the action (\ref{Riemann2}).

An essential advantage of the action (\ref{Riemann2}) with respect to  Einstein's action
is the fact that it is positive definite and has the nonzero value on the instanton solution
(it is proportional to the Euler signature of the space) in the exact analogy 
to the Yang--Mills case. Einsten's action of the gravity instanton is equal to zero 
(one can see immediately that the scalar curvature is absent at all points of the space due to 
$R_{\mu\nu} = 0$ equation, while the surface term, which must be accounted for in general
\cite{Gibbons}, turns out to be zero in this case \cite{Eguchi}. The zero action
for asymptotically flat spaces with a nontrivial metric meets troubles in quantum theory. 
In this case, the common flat vacuum is not marked out anyhow, which makes the perturbation
theory meaningless and contradicts the developed physical intuition. 

The theories with the action $S \sim R^2$ and $S \sim R_{\mu\nu}^2$ share 
this troublesome feature. The choice (\ref{Riemann2}) is also not optimal in our opinion.
 Any linear combination 
of the action (\ref{Riemann2}) with two terms   $\sim R^2$ and $\sim R_{\mu\nu}^2$
does also acquire nonzero positive value on the instanton solution. The most
attractive choice is
  \be
\label{Weyl2}
S \ =\ \frac 1\lambda \int d^4x \, \sqrt{-g} \, C_{\mu\nu\rho\sigma}  C^{\mu\nu\rho\sigma} \, ,
  \ee  
where $C_{\mu\nu\rho\sigma}$ is the Weyl tensor. The theory (\ref{Weyl2}) enjoys the local scale
invariance [the action is invariant under the transformation 
$g_{\mu\nu}(x) \to \lambda(x) g_{\mu\nu}(x)$] so that the action (\ref{Weyl2}) 
is marked out with respect to all other square curvature forms.

The action (\ref{Weyl2}) has an important additional advantage as 
compared to the action (\ref{Riemann2}) --- it reproduces itself when accounting 
for the perturbative quantum corrections. Under the choice (\ref{Riemann2}), counterterms would
arise which would not reduce to the initial form \cite{Fradkin}.

To avoid misunderstanding, note that the form $ R_{\mu\nu\rho\sigma}  R^{\mu\nu\rho\sigma}$
is independent of the forms $R^2$ and $R_{\mu\nu}^2$ in the full quantum theory. The 
Gauss--Bonnet identity tells that the linear combination
$R_{\mu\nu\rho\sigma}  R^{\mu\nu\rho\sigma} - 4 R_{\mu\nu} R^{\mu\nu} + R^2$ 
is a total derivative. It is of no account in the classical theory and the 
quantum perturbation theory, but in view of the topologically nontrivial solutions, the total
derivative in the Lagrangian cannot be neglected. (Remind the $\theta$--term problem in QCD. 
It is a total derivative, but its existence would lead to observable CP-nonconserving effects
in the strong interactions.)

It is time now to display difficulties of the theory under discussion. 
First of all, the equations of motion include the fourth derivatives of metric here. 
This leads to the ghost pole in the graviton propagator, which breaks unitarity. 
However, the ghost pole acquires an imaginary part under account of  the quantum 
corrections and drops out of the spectrum of physical asymptotic states \cite{LeeWick}.
It spoils the analytical properties of the scattering amplitudes, however, 
so that the theory is no longer causal. A clear understanding of this question 
is presently absent. One can say, e.g., that  noncausalities are restricted to the intervals
$\sim 10^{-33} {\rm cm}$ and do not contradict the observed macroscopic causality 
of the Nature.

I would like to speculate about another solution of this problem. We did see 
that the theory is rather similar to the Yang--Mills theory. The latter is known to include
the confinement phenomenon so that the quark and gluon poles appearing
 in the initial Lagrangian drop out completely as the poles of the physical scattering
amplitudes. We may guess that the same situation holds in our case, and the physical spectrum
 is presented by composite states with the masses of the Plank scale and the graviton, which
may be viewed as the Goldstone particle resulting from spontaneous breaking of conformal
symmetry. In this case, the causality of the real physical amplitudes may be restored.

Besides the asymptotic freedom and the presence of the instanton solutions, a strong evidence
for confinement is the nonrelativistic limit of conformal equations of motion. It is
 $\triangle \triangle \phi \sim \lambda \mu(\vec{r})$, where $\mu(\vec{r})$ is the mass density
and $\phi$ is the nonrelativistic gravity potential. The solution to this equation in the
 pointlike mass case is $\phi(\vec{r}) \sim \lambda m r$. The linearly rising potential, 
seemingly, leads to confinement of sources and, rather plausibly, to confinement of gauge
particles.

The most serious difficulty which the theory under consideration shares with all other theories
involving the spontaneous generation of the gravity constant is the appearance of the large
cosmological term $\Lambda_{\rm cosm.} \sim m_P^4$. The problem of the cosmological 
term is likely to be solved in the supersymmetric generalization of the action
(\ref{Weyl2}). The ${\cal N} =1$ superconformal gravity theory was constructed 
in Ref. \cite{Kaku}. It also has the property of the asymptotic freedom \cite{Fradkin2} 
so that the fundamental scale exists here where the conformal symmetry is broken and the 
dimensional gravity constant is induced. This theory is very complicated, and I am not able
 to answer the question whether the supersymmetry also breaks at this scale and the crucial
 question (which is connected with the first one) --- whether the cosmological term arises as a
 result of the symmetry breaking. 

I hope that it does not and that the conformal supergravity theory is an actual theory for
gravity interactions.

\vspace{1mm}

The author has benefited much from the discuttions with B.L. Ioffe, Ya.I. Kogan, 
and N.A. Voronov.

\end{document}